\documentclass[manuscript]{aastex}

\shorttitle{The IMF of Tr 14 & Tr 16}
\shortauthors{Hur et al.}
\begin{document}
\title{Distance and the Initial Mass Function of Young Open Clusters \\
in the $\eta$ Carina Nebula: Tr 14 and Tr 16}
\author{Hyeonoh Hur and Hwankyung Sung}
\affil{Department of Astronomy and Space Science, Sejong University, 98 Kunja-Dong, 
Kwangjin-gu, Seoul 143-747, Korea}
\email{hhur@sju.ac.kr,sungh@sejong.ac.kr}
\and
\author{Michael S. Bessell}
\affil{Research School of Astronomy and Astrophysics, Australian National University, MSO, Cotter Road, Weston, ACT 2611, Australia}
\email{bessell@mso.anu.edu.au}

\begin{abstract}
We present new $UBVI_c$ CCD photometry of the young open clusters Trumpler 14 (Tr 14) and Trumpler 16 (Tr 16) in the $\eta$ Carina nebula.
We also identify the optical counterpart of {\it Chandra} X-ray sources and Two Micron All Sky Survey point sources.
The members of the clusters were selected from the proper motion study, spectral types, reddening characteristics, and X-ray or near-IR excess emission.
An abnormal reddening law $R_{V,cl}=4.4\pm0.2$ was obtained for the stars in the $\eta$ Carina nebula using the 141 early-type stars
with high proper motion membership probability ($P_\mu\geqq70\%$).
We determined the distance to each cluster and conclude that
Tr 14 and Tr 16 have practically the same distance modulus of $V_0-M_V=12.3\pm0.2$ mag ($d=2.9\pm0.3$ kpc).
The slope of the initial mass function was determined to be $\Gamma=-1.3\pm0.1$ for Tr 14, $\Gamma=-1.3\pm0.1$ for Tr 16,
and $\Gamma=-1.4\pm0.1$ for all members in the observed region for the stars with $\log m\geqq0.2$.
We also estimated the age of the clusters to be about 1 -- 3 Myr from  the evolutionary stage of evolved stars and low-mass pre-main-sequence stars.
\end{abstract}

\section{Introduction}
The universality of the initial mass function (IMF) has been debated from the time that \citet{sa55} first introduced the idea of an IMF.
The massive part of the IMF can be represented by its slope from the inflection point at $\sim1M_{\sun}$ to the most massive stars
and is described by its slope ($\Gamma =\frac{d\log\xi}{d\log m}$), which is the so-called Salpeter-type IMF.
Several star forming regions show somewhat shallower slopes of $\Gamma$ in the massive part, particularly for the most massive star forming clusters,
even though we take the observational uncertainties in the photometric calibrations into account.
For example, the Arches cluster shows $\Gamma=-1.1\pm0.2$ \citep{esm09} and the core of NGC 3603 shows $\Gamma=-0.9\pm0.1$ \citep{sb04}.
However, the main reason for the difficulty in investigating the IMF of these clusters is that they are either highly reddened or very distant.

The $\eta$ Carina nebula (NGC 3372) is the largest nebula in the southern sky and is one of the most interesting regions in the Galaxy.
More than 60 O-type stars and several young open clusters are located in or near the nebula,
including Tr 14 and Tr 16 in the bright part of the nebula.
Moreover, $\eta$ Carinae, which is known as one of the most massive stars in the Galaxy, is near the center of Tr 16.
Although the Arches cluster and NGC 3603 appear to be more massive,
the young open clusters in the $\eta$ Carina nebula offer many advantages.
They are relatively close and less reddened, and are therefore able to be investigated in more detail.
Many previous investigators have studied these clusters.
\citet{wa73,wa95}, \citet{wa02}, \citet{le82}, \citet{mo88}, and \citet{mj93} presented spectral types of the stars in the region.
\citet{cu93} performed a proper motion study and presented membership probability of stars down to $V=16$ mag.
Previous studies include optical photometry \citep{va96,cu93,ta03,ca04},
near-IR (NIR) $JHK$ photometry \citep{sm87,ta03,as07,pr11}, or X-ray observations \citep{al08}.
\citet{mj93} obtained the slope of the IMF ($\Gamma=-1.3\pm0.2$) for Tr 14 and Tr 16 for $M\geqq6M_{\sun}$.
\citet{sm08} summarized the observational evidences of ongoing star formation activity in the $\eta$ Carina nebula.
Recently, \citet{tw11} (the {\it Chandra} Carina Complex Project -- CCCP) published a vast amount of information on the
stellar content and star formation activity in the $\eta$ Carina nebula region.

As these clusters are in the line of sight toward the tangential point of the Sagittarius spiral arm,
it has been questioned as to whether the young open clusters in the $\eta$ Carina nebula are at the same distance or not.
Tr 14 and Tr 16, in particular, have been at the center of this controversial question.
In addition, it is also argued that these two clusters exhibit an abnormal reddening law.
\citep{fe73,he76,fo78,th80,sm87,ta88,mj93,va96,ta03,ca04,as07}.
Most photometric studies have estimated the distances of these clusters to be 2.0--4.0 kpc.
Recently, \citet{sm06} obtained $2.35\pm0.5$ kpc for the distance to $\eta$ Carinae using the proper motion of the Homunculus Nebula around $\eta$ Carinae.

The aims of this study are (1) to judge whether the reddening law in the $\eta$ Carina nebula is abnormal or not,
(2) to determine the distance to the young open clusters in the nebula and decide whether they are at the same distance or not,
and (3) to obtain the IMF of the young open clusters in the nebula down to the inflection point $(\sim1M_{\sun})$
and determine its slope.
In Section~2 we present our new optical data and describe the data sets used in this work.
In Section~3, we determine the reddening law and distance modulus.
The membership selection criteria are presented in Section~4 where we construct the Herzsprung-Russell (H-R) diagram
and determine the age and masses of individual stars in the clusters. The IMF is derived in the same section.
The summary is presented in Section~5.

\section{Observation \label{obs}}
\subsection{Optical data\label{optic}}
$UBVI_c$ observations were performed on 1997 March 2, June 23, 1999 February 5 and 6 using the 1m telescope at Siding Spring Observatory with a SITe 2K CCD camera.
The field of view was $20\farcm5\times20\farcm5$.
We obtained two short and four long exposure images for each filter.
The average seeing was $\sim2\arcsec$ during all observing runs.
The total observed area covered $22\farcm4\times20\farcm8$ of the nebula.

We performed point-spread-function (PSF) photometry using the DAOPHOT package in IRAF
\footnote{Image Reduction and Analysis Facility is developed and distributed by the National Optical Astronomy Observatories which is operated by the Association of Universities for Research in Astronomy under operative agreement with the National Science Foundation.}.
An aperture correction was applied to produce the equivalent magnitude for a $7\arcsec$ radius.
We transformed our observed data to the SAAO standard system using SAAO E5 and E7 regions \citep{su00}.
The photometric data down to $V=19$ mag are presented in Table~\ref{tab1}.
\placetable{tab1}

We compared our photometry with previous studies, and the results are presented in Table~\ref{tab2}.
Three studies \citep{va96,ta03,ca04} were based on CCD PSF photometry while \citet{mj93} performed simple aperture photometry. Others were based on photoelectric photometry.
In the comparison, we have excluded stars that deviated by more than $2.5\sigma$ from the mean to avoid the inclusion of variables or optical doubles.
Our photometry is in good agreement with previous photoelectric photometry and CCD photometry by \citet{va96}.
But although we find no significant difference in the photometric zero point in $V$,
we find a large scatter in the comparison with the photoelectric data and simple aperture photometric data of \citet{mj93}.
Such a large scatter may be caused by difficulty in correcting for the sky background due to spatially varying nebulosity.
There are no systematic differences in color between our photometry and previous photometry except with \citet{ta03} and \citet{ca04}.

\begin {figure}
\plotone{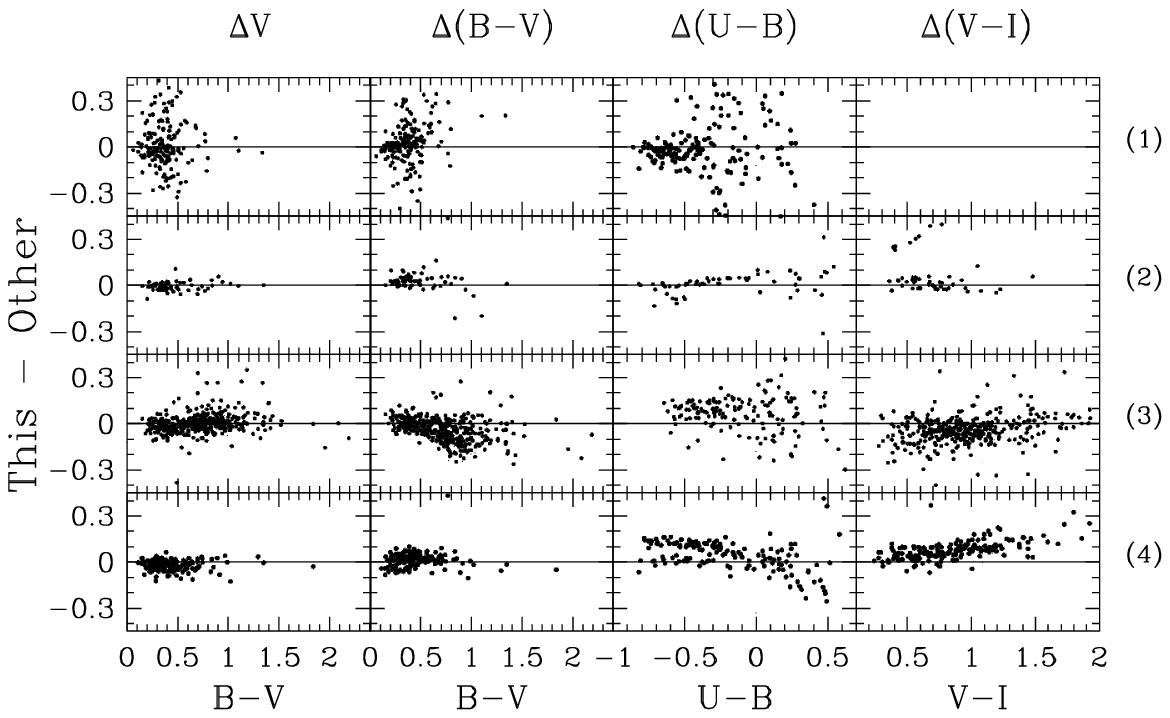}
\caption{Comparison of our photometry with four previous sets of CCD photometry. 
Each panel from upper to lower shows the difference between us and 
(1) \citet{mj93}, (2) \citet{va96}, (3) \citet{ta03}, and (4) \citet{ca04}, respectively.
\label{delta}}
\end{figure}

The comparison with previous CCD photometry is shown in Figure~\ref{delta}.
The top panel shows the comparison with the simple aperture photometric data of \citet{mj93}.  This shows the largest scatter, as mentioned above.
The second panel shows the comparison with the CCD photometry of \citet{va96}.
There is no significant systematic difference with their data except in $(V-I)$ for some stars around Tr 14.
Most of these stars are brighter than $V < 14.5$ mag and located in a small area ($-3\farcm5 < \Delta$ R.A. $< -7\farcm5$, $7\farcm0 < \Delta$ decl. $< 10\farcm5$).
In the third and the last panels we show the comparison with CCD photometry by \citet{ta03} and \citet{ca04}.
Their data show large differences in the photometric zero points as well as a large scatter.
\placetable{tab2}
\placefigure{delta}

\subsection{X-Ray data and 2MASS data\label{x_2mass}}
\begin {figure}
\plotone{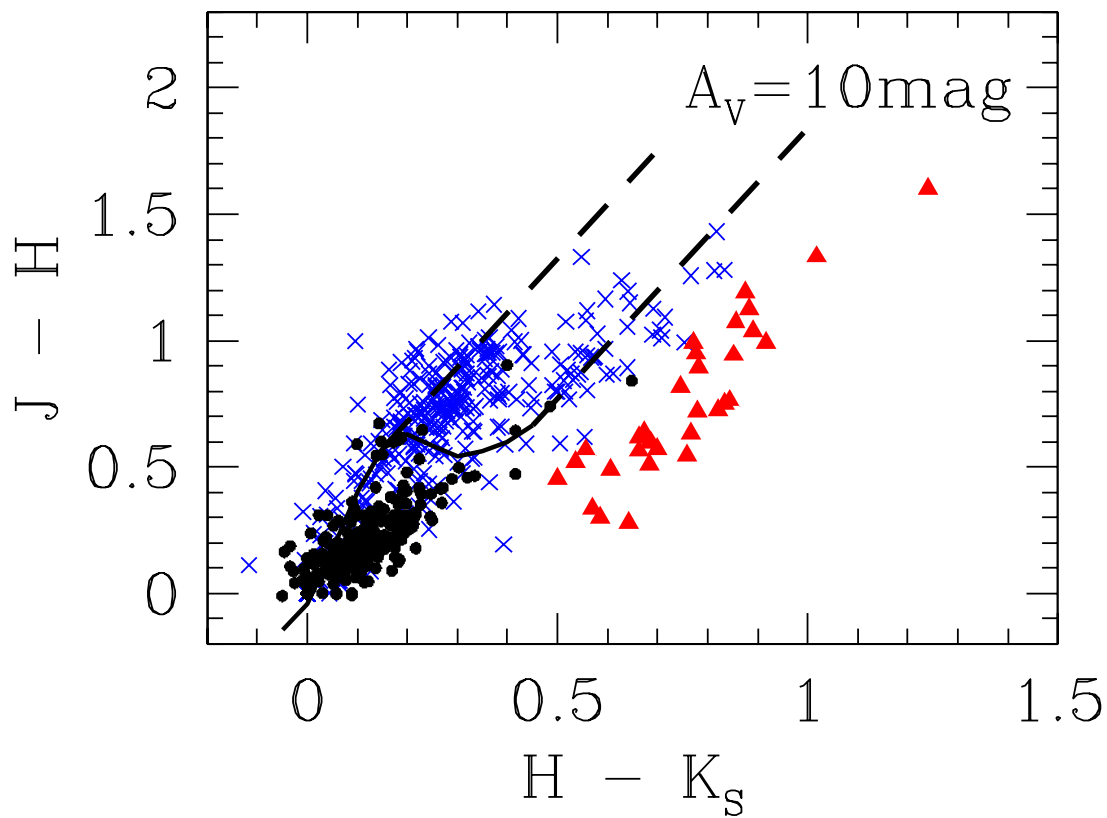}
\caption{
2MASS color--color diagram of stars with $\epsilon(J-H:H-K_S)<0.1$ mag.
The solid and dashed lines are the MS relation \protect\citep{su08} and the reddening vector \protect\citep{fi99} for $A_V=10$ mag.
A triangle (red), filled circle (black), and cross (blue) represent NIR excess stars, proper motion members ($ P_{\mu} {\protect\geqq} 70\%$), and X-ray emission stars, respectively.
\label{jhk}}
\end{figure}

CCD coordinates have been transformed to the equatorial coordinate system using the Two Micron All Sky Survey (2MASS) point-source catalog \citep{sk06}.
A total of 4615 out of 6235 optical sources have been identified with 2MASS point sources within a search radius of $1\arcsec$.
Figure~\ref{jhk} shows the ($J-H$) versus ($H-K_S$) color--color diagram for the stars with $\epsilon(J-H:H-K_S) (\equiv\sqrt{{\epsilon_J}^2+2{\epsilon_H}^2+{\epsilon_{K_S}}^2}) <0.1$ mag.
The main-sequence (MS) relation of \citet{su08} and the reddening law of \citet{fi99} have been adopted and used in the figure.
Thirty-four stars were found to have excess emission in $(H-K_S)$ color and are classified as NIR excess stars.
\placefigure{jhk}
X-ray observations are the most powerful tool for selecting PMS members \citep{su04}.
The {\it Chandra} X-ray point source catalog published by \citet{al08} and the CCCP X-ray source catalog published by \citet{br11} were used in selecting the X-ray emission stars.
\citet{al08} also published X-ray sources in the $17\arcmin\times17\arcmin$ area centered on R.A.$=10\fh44\fm47\fs93$ and decl.$=-59\degr43\fm54\fs21$, which covers the entire region of Tr 16.
The CCCP provided a catalog of 14369 X-ray sources in a 1.42 deg$^2$ area including the region observed by \citet{al08}.
There were 3338 CCCP sources in our field of view.
Both X-ray catalogs supplied the 2MASS counterpart of X-ray sources.
First, we used their identification and found 964 and 373 X-ray emission stars in the CCCP catalog and \citet{al08}, respectively.
Next, we searched for optical counterparts of X-ray sources without a 2MASS ID using a search radius of $1\farcs2$.
If two or more stars were found within the radius, we assigned the closest star as the optical counterpart of the X-ray source.
We identified 192 and 96 stars as X-ray emission stars from the CCCP catalog and \citet{al08}, respectively.
As a result, a total of 1184 stars were selected as X-ray emission stars. Among them, 441 stars were identified as X-ray sources from  both X-ray catalogs.

\section{Reddening and Distance \label{RD}}
\subsection{Photometric Diagrams \label{phd}}
\begin {figure}
\plotone{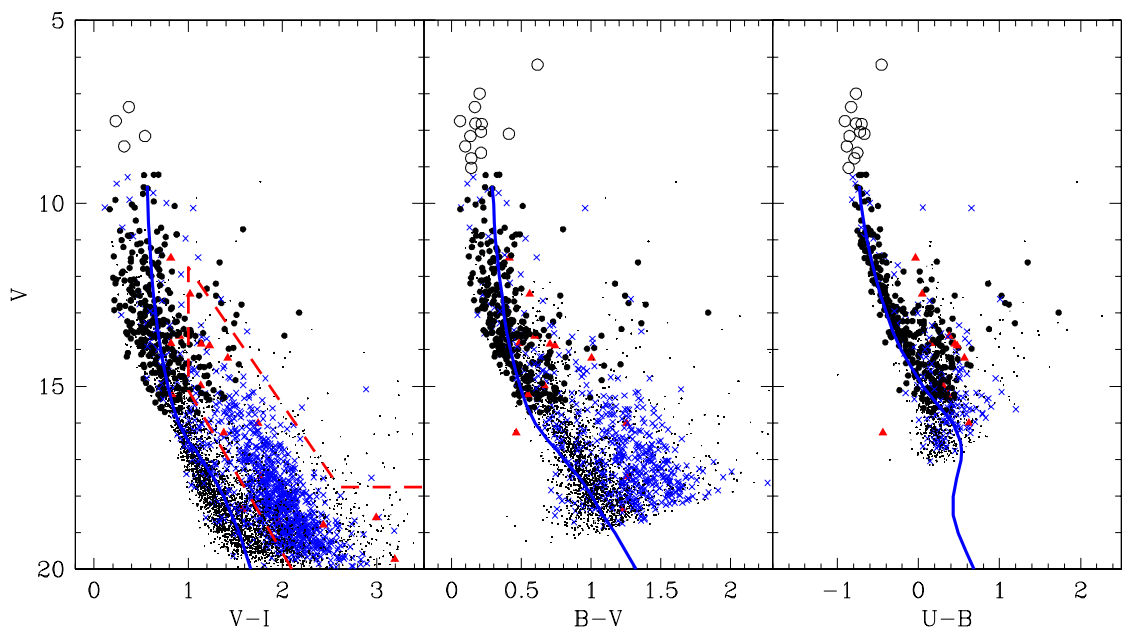}
\caption{Color--magnitude diagrams of stars with $\epsilon\protect\leqq0.1$ mag.
An open circle, filled circle, cross (blue), and triangle (red) represent data from photoelectric photometry, proper motion member, X-ray emission star, and NIR excess star, respectively.
The stars with multiple membership criteria are marked with the symbol according to their primary membership criterion.
The dashed and solid lines represent, respectively, the PMS locus and the ZAMS relation with $V_0 - M_V = 12.3$ mag and $E(B-V)=0.61$ mag. 
\label{cmd}}
\end{figure}

Figure~\ref{cmd} shows the color--magnitude diagrams (CMDs) of stars with $\epsilon<0.1$ mag in the observed region.
Thick lines represent the reddened zero-age main-sequence (ZAMS) relation from \citet{su97} for $E(B-V)=0.61$ mag.
The stars brighter than $V = 9$ mag were saturated in our observations.
We therefore used the mean magnitudes and colors from the SIMBAD\footnote{http://simbad.u-strassbg.fr/simbad/}
database for these stars and they are marked as open circles in the figures.

The stars with high membership probability ($P_\mu\geqq70\%$) from the proper motion study \citep{cu93}
were classified as proper motion members (see \S~\ref{mem}) and are marked as filled circles.
X-ray emission stars and NIR excess stars are marked as crosses and triangles, respectively.
Most of the proper motion members are scattered around the reddened ZAMS line.
The relatively large scatter implies a wide range of reddening among the cluster members.
There are several red stars ($V-I > 1.0$ and $V < 3.75 (V-I) + 8.0$) with a high proper motion membership probability.
Seven stars (ID 1817, 2599, 2639, 2754, 2830, 3337, and 3911) seem to be foreground late-type stars with similar proper motion vectors by chance to the clusters.
Four stars (ID 2022, 2033, 4211, and 4238) are near the boundary of bright nebula; they are probably highly obscured ($E(B-V)> 0.9$ mag) early-type stars.
Six additional stars (ID 677, 849, 1027, 1374, 2462, and 4044) have similar photometric characteristics as the four stars mentioned above.
The star ID 849 is an optical double, and shows a variability of $\Delta V\sim0.25$ mag from four observations.
In addition, the star ID 2572 ($V=16.265, V-I=1.375, B-V=0.461, U-B=-0.440$) shows very strong UV excess with a NIR excess; this star appears to be a classical T Tauri star with active accretion.

Most X-ray emission stars are fainter than $V=15$ mag and redder than the reddened ZAMS line by about 0.6 mag.
They constitute a well-defined sequence of PMS stars in the clusters.
However, a non-negligible number of X-ray emission stars can also be found around the reddened ZAMS line in the ($V$, $V-I$) diagram.
They seem to be X-ray active field stars in front of the $\eta$ Carina nebula because their $V-I$ color is not affected by the UV excess and
late-type stars retain their X-ray activity for a long time.
There are several X-ray emission stars redder than most PMS stars with X-ray emission; 
two of them (ID 2022 and 4211), as already mentioned, are highly reddened early-type stars with high proper motion membership probability. The star ID 2412 seems to be a foreground late-type star because its position in the ($U-B$,$B-V$) diagram is close to the unreddened ZAMS line.
The X-ray emission star ID 1836 seems to be a late-type member with a strong UV excess.
The other three stars (ID 3729, 4081, and 4212) are very difficult to justify as members from optical data alone; 
they seem to be highly reddened early-type stars in the ($J-H, H-K_s$) diagram and, in addition, are in the dark lane to the south of Tr 16.
The membership of these presumably highly reddened stars, including the four stars mentioned above, is uncertain,
and they are for the time being, not included in the member list.
\placefigure{cmd}

\begin {figure}
\plotone{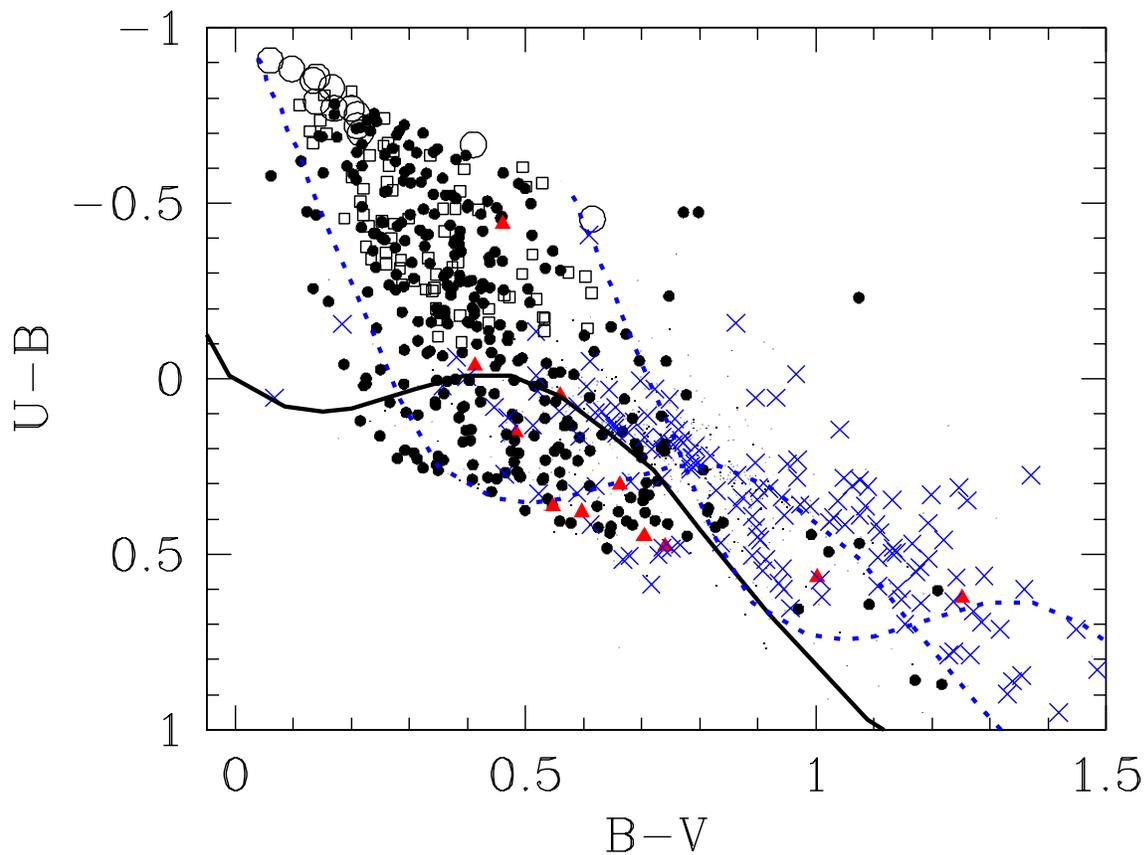}
\caption{($U-B$) vs. ($B-V$) color--color diagram of the stars with $V\protect\leqq17$ mag.
Open squares represent members selected from the reddening law (see Section~\ref{mem}).
The solid line is the unreddened MS relation, while dotted lines are reddened MS relations with $E(B-V)=0.36$ and 0.9 mag.
Other symbols are the same as Figure~\ref{cmd}. The large open circle near $E(B-V)=0.9$ mag represents $\eta$ Carinae. 
\label{ccd}}
\end{figure}

\begin {figure}
\plotone{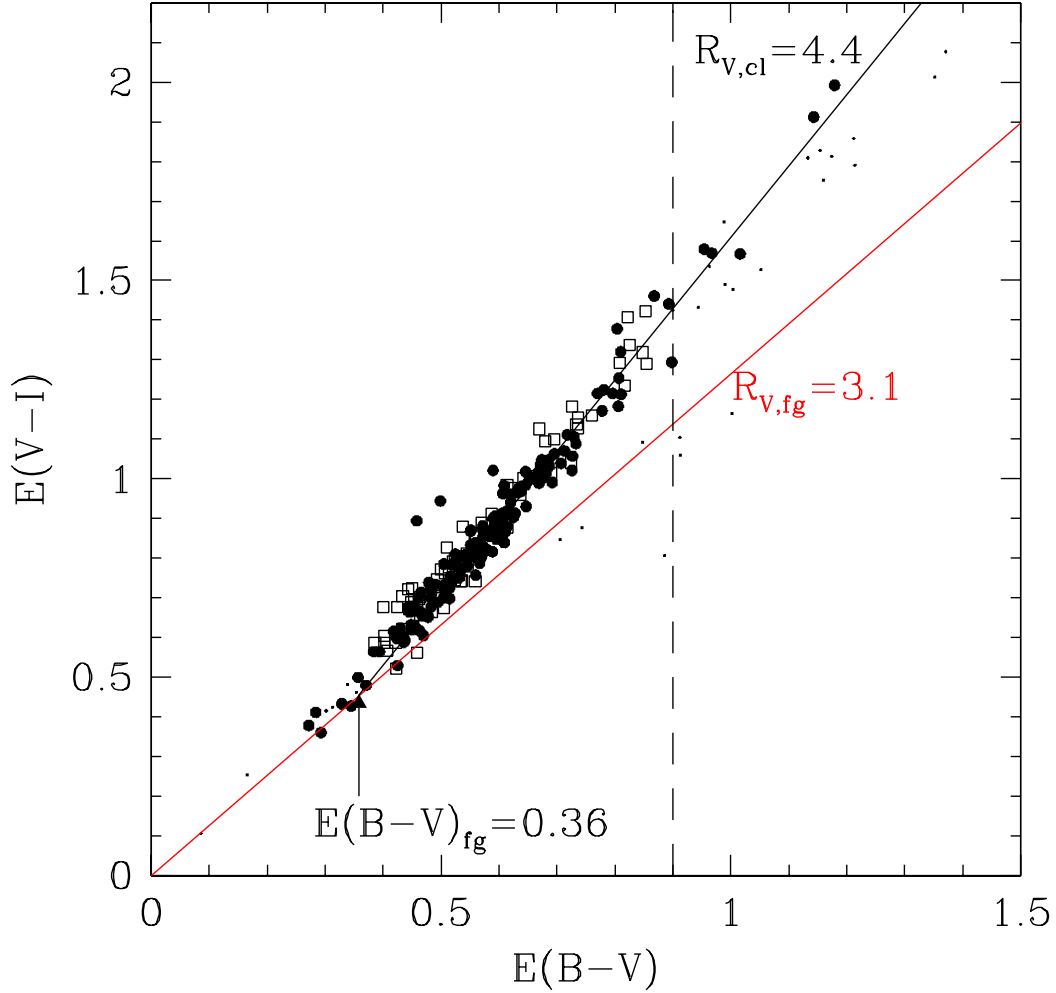}
\caption{Reddening law of early-type stars in the $\eta$ Carina nebula.
Solid lines correspond to $R_V=3.1$ and $R_{V,cl}=4.4$ for $E(B-V) \protect\geqq E(B-V)_{fg}$, and dashed line corresponds to $E(B-V)=0.9$.
Filled circles, open squares, and dots represent the stars with $P_\mu \protect\geqq 70\%$, the members selected from the reddening law (see Section~\ref{mem}),
and the early-type stars with $P_\mu < 70\%$ or without proper motion membership probability. 
\label{rv}}
\end{figure}

Figure~\ref{ccd} shows the ($U-B$, $B-V$) color--color diagram.
Early-type members without proper motion membership probability were selected from the locus of early-type members
in Figure~\ref{rv} (see Section~\ref{RL}) and are marked as open squares in Figure~\ref{ccd} and \ref{rv}.
The reddening of proper motion members and bright members is between $E(B-V)$  = 0.36 and 0.9.
There are several stars with $E(B-V)<E(B-V)_{fg}=0.36$ mag and with high proper motion membership probability ($P_{\mu}\geqq 70\%$)
including six early-type stars (ID 2154, 2377, 2475, 2693, 3223, and 4156).
Were they actual members of the $\eta$ Carina nebula, they should be on the blue part of the reddened ZAMS whereas they lie in the middle of the MS band in the $(V,U-B)$ CMD. 
This implies that they are more likely foreground stars in the spiral arm between $\eta$ Carinae and the Sun (see \S~\ref{RL}).
There are also several proper motion members in the red part of the $E(B-V)=0.9$ line.
Blue stars with $Q \equiv (U-B)-0.72(B-V) \leqq -0.5$ mag are highly reddened early-type stars,
while most red stars near the $E(B-V)=0.9$ line are foreground late-type stars.
In addition, there are many X-ray emission stars between $(B-V)=$ 0.9 -- 1.3 mag and $(U-B)=$ 0.0 -- 0.8 mag.
These are PMS stars with UV excess \citep{su97} implying that the PMS stars in Tr 16 are actively accreting.
\placefigure{ccd}

\subsection{Reddening law\label{RL}}
It has long been debated whether or not Tr 14 and Tr 16 are at the same distance.
The reddening law may play a key role in this debate.
The total extinction in the $V$ band can be determined using the relation $A_V={R_V\times E(B-V)}$.
The so-called normal reddening law of $R_V=3.1\pm0.2$ is consistently obtained for stars in the solar neighborhood \citep{gu89,li11},
while many previous studies noted that the total-to-selective extinction ratio $R_V$ toward the $\eta$ Carina region is anomalously higher
\citep{fe73,he76,fo78,th80,sm87,ta88,va96,sm02}.
\citet{va96} used the membership probability ($P_\mu$) \citep{cu93} as the membership selection criterion
and determined $E(B-V)_{fg}=0.33$ mag and $R_{V,cl}=4.7\pm0.6$ for Tr 14
[$A_V = 3.1\times E(B-V)_{fg} + R_{V,cl}\times [E(B-V)-E(B-V)_{fg}]$, see also \citet{fo78}].

We have determined the reddening law toward the $\eta$ Carina nebula using the relation between
$R_V$ and color excess ratios, $R_V=2.45{E(V-I)\over E(B-V)}$ \citep{gu89}.
The $E(B-V)$  reddening of individual early-type stars is determined in the $(U-B,B-V)$ diagram using the slope of the reddening vector ${E(U-B)\over E(B-V)}=0.72$.
The reddening in $(V-I)$, $E(V-I)$, is the difference between the observed and the intrinsic $(V-I)$. The intrinsic $(V-I)$ is determined from the relation between $(B-V)_0$ and $(V-I)_0$ for MS stars (see Appendix of \citet{su99}). 
We selected early-type stars to be those with $V<15.5$ mag and $Q= (U-B) -0.72(B-V) \leqq -0.5$ mag or with $V<15.5$ mag, $(U-B) <-0.1$ mag and $(B-V) <0.45$ mag.
The early-type membership selection criteria are $P_\mu \geqq 70\%$ and $0.36\leqq E(B-V)\leqq0.9$ mag.
Stars with $E(B-V)>0.9$ mag were excluded to avoid contamination by PMS stars with strong UV excess or background early-type stars.
A total of 141 early-type members were selected and used to determine the reddening law and distance modulus of the $\eta$ Carina nebula.
Figure~\ref{rv} shows the relation between $E(B-V)$ and $E(V-I)$.
We determined the color excess ratio due to the intracluster dust to be $E(V-I)_{cl}\over E(B-V)_{cl}$
($\equiv {E(V-I) - E(V-I)_{fg}\over E(B-V) - E(B-V)_{fg}}$) = $1.80 \pm 0.10$
and $E(B-V)_{fg}=0.36\pm0.04$ mag using least squares.
The color excess ratio ${E(V-I)\over E(B-V)}$ for a normal reddening law is 1.25 \citep{de78}.
The total-to-selective extinction ratio in the $\eta$ Carina region ($R_{V,cl}$) is, therefore, $4.4 \pm 0.2$.

If the six early-type stars with $E(B-V)<0.36$ mag are included,
the resultant foreground reddening $E(B-V)_{fg}$ and reddening law $R_V$ are $0.25\pm0.09$ and $4.0\pm0.25$, respectively.
Although we could derive the same distance modulus in the case of a smaller $R_V$,
the reddened ZAMS lines are not well fitted to the distribution of early-type members in Figure~\ref{cmd}.
This fact supports the suggestion that these less reddened early-type stars are foreground stars in the Sagittarius spiral arm 
between $V_0-M_V =$ 11.3 --11.8 mag (see Section~\ref{phd}).

\citet{va96} also applied the same method to determine the reddening law of the nebula.
Although there are small differences in the membership selection criteria, their result agrees well with ours.
The small difference between our result and theirs appears to be caused by the number of stars used in the determination of the reddening law (53 stars were used in \citet{va96}, while 141 stars were used in this work).
\placefigure{rv}

\begin {figure}
\plotone{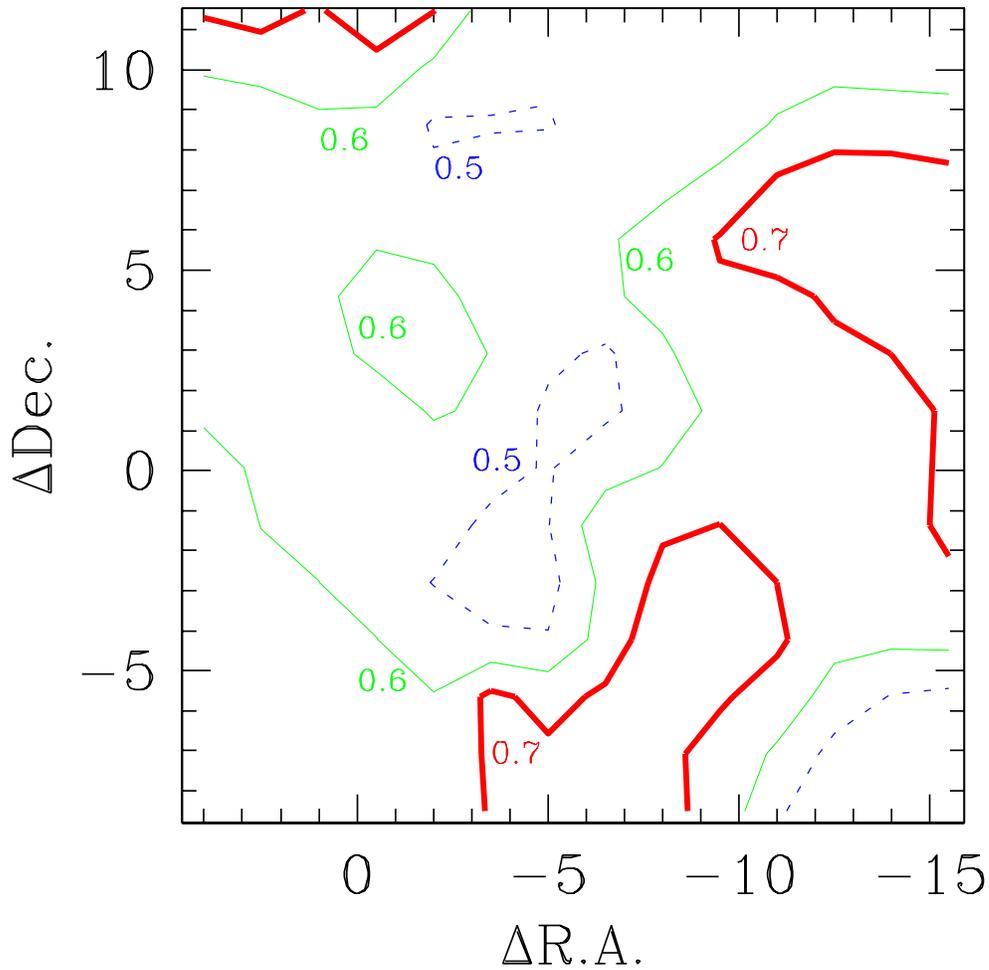}
\caption{Smoothed reddening map of the observed region.
The lines represent $E(B-V)=0.5$ (dashed line), 0.6 (thin solid line), and 0.7 (thick solid line), respectively.
$\Delta$ R.A. and $\Delta$ decl. represent the angular distance in minutes of arc from $\eta$ Carinae (R.A.$=10\fh 45\fm 3\fs591$, decl= $-59\arcdeg 41\arcmin 4\farcs26 $, J2000.0).
\label{redmap}}
\end{figure}

As there is no reliable photometric way to estimate the reddening of PMS stars,
we have to rely on the reddening map derived using the early-type stars.
Figure~\ref{redmap} shows the reddening map.
The dashed line represents $E(B-V)=0.5$ mag and the solid lines represent $E(B-V)= 0.6$ and $0.7$ mag, respectively, from thin to thick line.
The spatial variation of $E(B-V)$ in Figure~\ref{redmap} is well matched by the reddening map of \citet{fe73} and the extinction map derived by \citet{sm87}.
The highly reddened region is to the west of Tr 14 ($\Delta $R.A.$>-10\farcm0$ and $4\farcm0<\Delta $decl.$<8\farcm0$, 
where $\Delta$R.A. and $\Delta$decl. represent the angular distance in R.A. and Dec. from $\eta$ Carinae) and is well matched by the CO emission map of \citet{br03}.
\placefigure{redmap}

\subsection{Distance\label{distance}}
\begin {figure}
\plotone{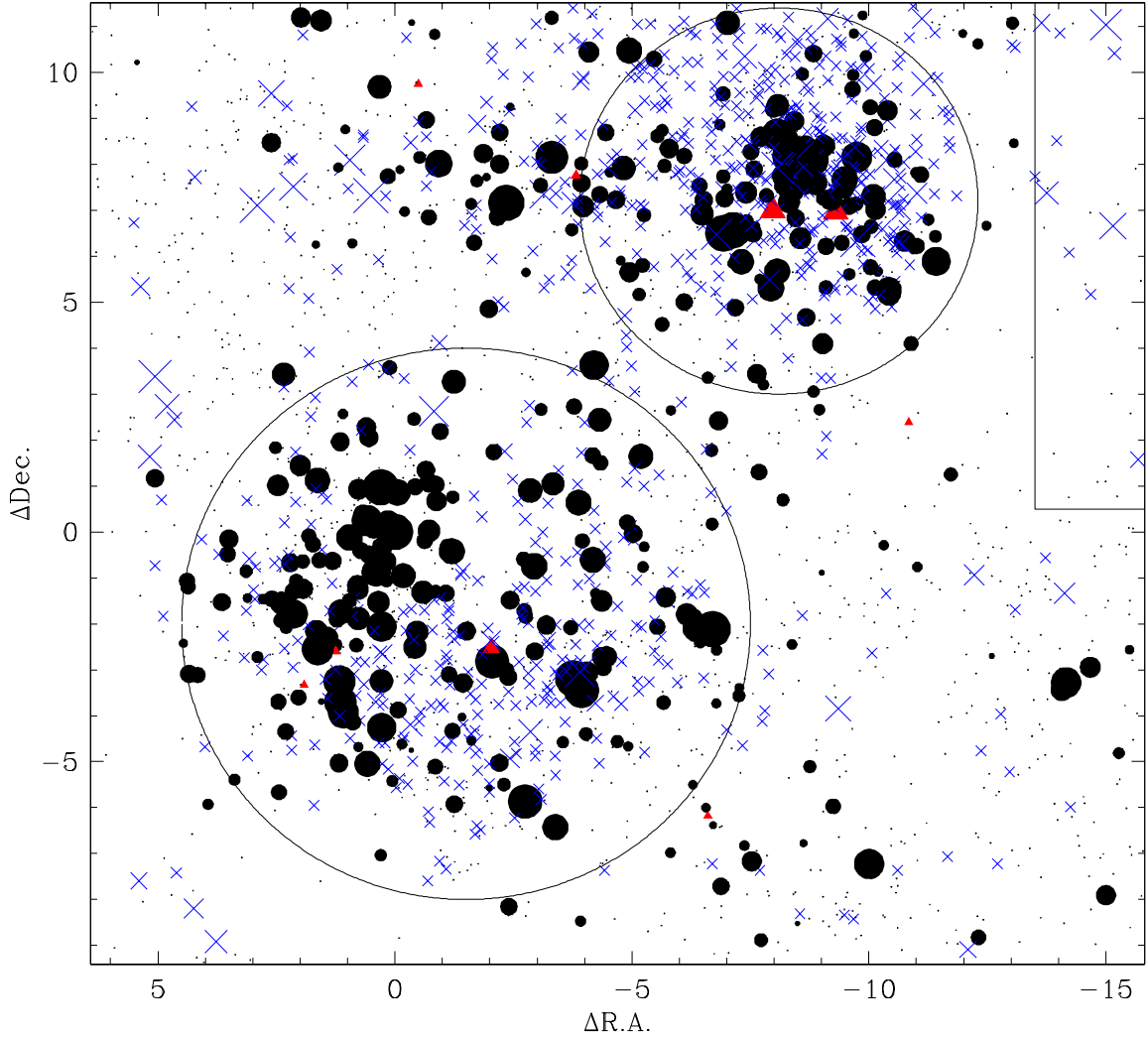}
\caption{
Spatial distribution of member and candidate stars.
Filled circles represent members from the proper motion study ($P_\mu\protect\geqq70\%$), and from the abnormal reddening law in Figure~\ref{rv}.
$\Delta$ R.A. and $\Delta$ decl. represent the angular distance in minutes of arc from $\eta$ Carinae.
The members with X-ray emission or NIR excess are marked with a cross (blue) or a triangle (red).
The stars with multiple membership criteria are marked by their primary membership criterion mentioned above.
The size of symbols of members is proportional to the $V$ magnitude of the stars.
Small dots represent the candidate stars.
The large circles represent the limit of Tr 14 ($\Delta$R.A.$=-8\farcm1$, $\Delta$decl.$=7\farcm2$, radius = $4\farcm2$)
and Tr 16 ($\Delta$R.A.$=-1\farcm5$, $\Delta$decl.$=-2\farcm0$, radius = $6\farcm0$).
The rectangle to the west of Tr 14 represents the comparison field region used to remove the foreground contamination from the IMF (see Sections~\ref{distance} and~\ref{IMF}).
\label{plot}}
\end{figure}

In order to derive the distance modulus of each cluster, we should first determine the membership containing radius of each cluster.
However, as it is still uncertain whether Tr 14, Tr 16, and Cr 232 are dynamically independent clusters, it is therefore very difficult to set the boundaries of each cluster.
We decided to set the centers and boundaries of each cluster so as to include most of proper motion members ($P_\mu\geqq70\%$).
The center and radius of each cluster are $\Delta$ R.A.$=-8\farcm1$, $\Delta$ decl.$=7\farcm2$, and radius = $4\farcm2$ for Tr 14
and $\Delta$ R.A.$=-1\farcm5$, $\Delta$ decl.$=-2\farcm0$, and radius = $6\farcm0$ for Tr 16 (see Figure~\ref{plot}).
\placefigure{plot}

\begin {figure}
\epsscale{.90}
\plotone{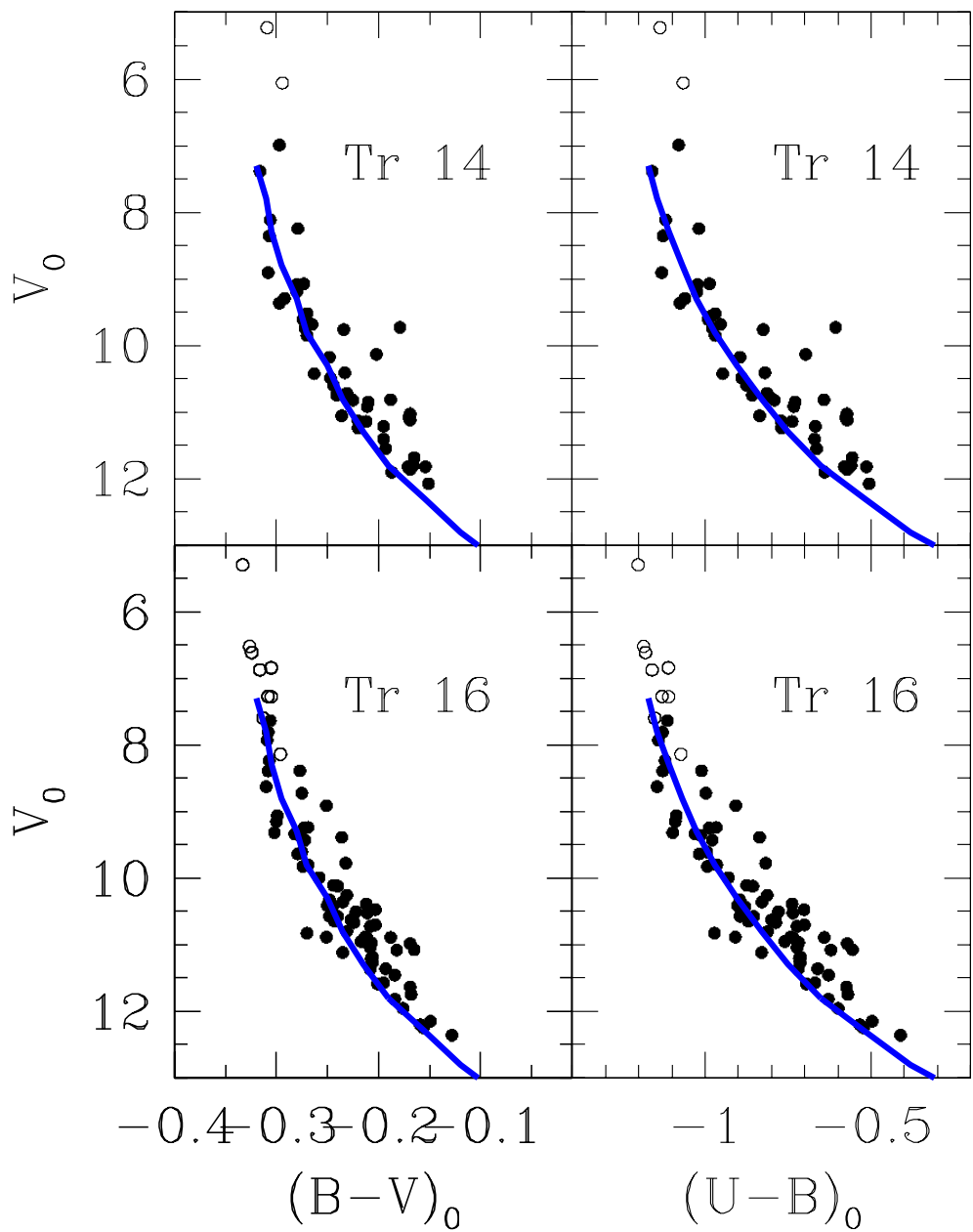}
\caption{
ZAMS fitting. 
Open symbols represent the known binaries from \citet{ga08}, \citet{ma09}, \citet{ne04}, and \citet{ra01,ra09}.
From the distribution of distance moduli of individual member stars,
we obtain the same distance modulus ($V_0-M_V=12.3$ mag) both for Tr 14 and Tr 16.
The thick line represents the ZAMS relation shifted by the adopted distance modulus. 
\label{dm}}
\end{figure}

Although the proper motion study and spectral classification studies can supply good membership criteria,
it is still debated whether Tr 14 and Tr 16 are at the same distance or not.
Two recent optical studies \citep{ta03,ca04} arrived at different conclusions.
\citet{ca04} determined the reddening law in three different ways and adopted an average value.
They adopted $R_V=3.48$ and $4.16$ for Tr 14 and Tr 16, respectively, and obtained a different distance modulus for each cluster
($V_0-M_V=12.3\pm0.2$ mag for Tr 14 and $13.0\pm0.3$ mag for Tr 16).
However, they did not take into account the difference, in star-forming regions, of the reddening law  between the general interstellar medium and the intracluster medium.
In addition, the number of members used in the determination of the reddening law was very small (10 stars for Tr 14 and 14 stars for Tr 16).
On the other hand, \citet{ta03} obtained $V_0-M_V=12.1$ mag using $A_V=1.39E(V-J)$ to avoid the anomalous reddening law at optical wavelengths
and concluded that Tr 14 and Tr 16 are at the same distance.
\citet{va96} determined $V_0-M_V=12.5\pm0.2$ mag for Tr 14 by applying the abnormal reddening law they determined (see Section~\ref{RL}).
Recently, \citet{sm06} determined the distance to $\eta$ Carinae as $d=2.35\pm0.05$ kpc $(V_0-M_V=11.85\pm0.05$ mag)
from the proper motion of the Homunculus Nebula ejected from $\eta$ Carinae.
\placefigure{dm}

We calculated the total extinction in the $V$ band of individual early-type members using the reddening-corrected colors
and the adopted ZAMS relation \citep{su97}, and constructed the distribution of distance moduli for each cluster.
We find the same distance modulus for Tr 14 and Tr 16 ($V_0-M_V=12.3\pm0.2$ mag, i.e. $d=2.9\pm0.3$ kpc) and conclude that Tr 14 and Tr 16 are at the same distance within the observational errors.
Figure~\ref{dm} shows the reddening corrected CMDs superimposed with the ZAMS relation shifted by the adopted distance modulus.
Although it is very difficult to estimate the error in the ZAMS fitting, we can expect an error of at least 0.1 mag.
In addition, the ZAMS relation itself may be uncertain by about 0.1 mag.
The error in the distance modulus determination is, therefore, about 0.2 mag.
Our distance modulus is in good agreement with that by \citet{ta03} and \citet{va96}, 
but disagrees with that by \citet{ca04} who estimated different distances to Tr 14 and Tr 16.
This difference is caused by the applied reddening correction, i.e. the difference in $A_V$ of individual early-type stars.

There is a non-negligible difference in the derived distance between the two methods,
the proper motion of the expanding nebula ejected from $\eta$ Carinae (the Homunculus Nebula) and the photometric analysis of stars in the nebula.
\citet{ah93}, \citet{me99}, and \citet{sm02, sm06} performed spectroscopic analyses of the Homunculus Nebula and determined $2.2\pm0.2$ kpc, $2.3\pm0.3$ kpc, and $2.35\pm0.05$ kpc, respectively.
These are quite consistent each other within the expected error.
On the other hand, the distance moduli determined by \citet{va96} and \citet{ta03} are comparable with our result of $d=2.9\pm0.3$ kpc.
Because these latter two studies and our work incorporate the abnormal reddening law toward the $\eta$ Carina nebula,
the consistency of the distance moduli from the photometric studies suggests that our distance determination is realistic.
Furthermore, the possibility cannot be ruled out that the distance from the proper motion studies of the ejected nebula might be affected by the assumption of a constant expanding velocity and symmetrical shape of the Homunculus Nebula.
If the expansion velocity of the ejected material varied after the initial ejection, or if the Homunculus Nebula is asymmetric in shape,
the proper motion of the expanding nebula may be underestimated.
In addition, as $\eta$ Carinae is known to be a binary system \citep{dcl97},
the influence of the companion star on the expansion of the Homunculus Nebula should be taken into account.
Therefore, although our distance modulus is somewhat larger than the distance determined from the proper motion of the ejected nebula,
our value reasonably explains the location of the MS band.

Recently \citet{ar11} studied the proper motion of the Homunculus Nebula and determined the epoch of the eruption of $\eta$ Carinae.
They assumed the distance to $\eta$ Carinae as 2.3 kpc. 
Although most of parameters (projected velocities, orientation angles, and projected distances from $\eta$ Carinae) may be changed 
if we adopt the distance to $\eta$ Carinae to be 2.9 kpc, 
the epoch of eruption is nearly the same because the projected velocity of the nebula generally dominates the radial velocity.  

\section{H-R diagram \label{HR}}
\subsection{Membership selection\label{mem}}
The reddening law was used as the primary membership selection criterion for early-type stars.
In Figure~\ref{rv}, most of the early-type stars with $P_\mu\geqq70\%$ lie along the line of $R_{V,cl}=4.4$.
The lower boundary of membership selection was set to be $\Delta E(B-V)=0.06$ mag at a given $E(V-I)$.
We classified stars above the lower boundary and with $0.36\leqq E(B-V)\leqq0.9$ mag as early-type members.
Early-type members selected using the reddening law and without proper motion membership probability are marked as open squares in Figure~\ref{rv}.
From this criterion, some stars with low membership probability ($20\%\leqq P_\mu<70\%$) were also selected as members
because nearly face-on binary systems may have a low proper motion membership probability.
For instance, HD 93250 (O3.5V((f+)), $P_\mu=24\%$), one of the earliest type stars in the $\eta$ Carina region, was selected as a member.
The star shows unusually strong X-ray activity \citep{ev04,na11} and is a suspected binary system.
But it is still not confirmed as a binary star due to a lack of radial velocity variations \citep{ra09}.  
For early-type members selected using the reddening criterion, the total extinction $A_V$ was calculated individually using 
$A_V = 3.1\times E(B-V)_{fg} + R_{V,cl}\times [E(B-V)-E(B-V)_{fg}]$ \citep{fo78}.

For intermediate or low-mass PMS stars, $E(B-V)$ was estimated from the reddening map of Figure~\ref{redmap}.
Stars with reddening-corrected colors within $\vert\Delta(V-I)_0\vert\leqq0.15$ mag and $\vert\Delta(B-V)_0\vert\leqq0.15$ mag from the ZAMS relation,
and with membership selection criteria (X-ray emission, $P_\mu\geqq70\%$ or NIR excess) were
classified as members and stars with $20\%\leqq P_\mu<70\%$ were classified as candidates.
To select the PMS members, we set the PMS locus in Figure~\ref{cmd} to include most of the X-ray emission stars \citep{su08}.
The thick dashed line in Figure~\ref{cmd} represents the PMS locus.
We classified stars satisfying the above membership selection criteria (i.e. X-ray emission, $P_\mu\geqq70\%$ or NIR excess and falling within the PMS locus) as members.
The stars in the PMS locus without membership criteria were classified as PMS candidates.
The spatial distribution of members is shown in Figure~\ref{plot}.

\subsection{Color--$T_{eff}$ Relation and Bolometric Correction}
In order to construct the H-R diagram, we transformed the reddening-corrected colors, spectral type, and $M_V$ to effective temperature and bolometric magnitude.
The temperature of the O2 supergiant HD 93129A (O2If$^*$) is somewhat uncertain.  
\citet{pu96} determined the temperature of HD 93129A to be 50,500 K, while \citet{re04} estimated it to be 42,500 K. 
However, a temperature of 42,500 K is much too low for spectral type O2. 
As there are no confirmed O2If$^*$ stars except HD 93129A in the Galaxy, we referred to the temperature of O2If$^*$ stars in the LMC. 
\citet{do11} determined the effective temperature of two O2If$^*$ to be about 50,000 K using \ion{N}{4} and \ion{N}{5} lines, instead of He lines. 
We, therefore, assumed $T_{eff}$ = 50,000 K for HD 93129A.
For O3 to O8 stars, we applied the spectral-type--temperature relation and the bolometric correction from \citet{ma05}.
For O9 to $(V-I)_0<1.6$ mag stars, we adopted the color--temperature and temperature--bolometric correction relations from \citet{be98}.
For $(V-I)_0 <0$ mag stars, $(B-V)_0$ and $(U-B)_0$ were used as the temperature indicators.
For $0\leqq(V-I)_0\leqq1.6$ mag stars, $(V-I)_0$ was used as the temperature indicator.
For $(V-I)_0>1.6$ mag stars, we have adopted the empirical color--temperature relation and bolometric correction scale of \citet{be91}.
\citet{da97} estimated the total luminosity of $\eta$ Carinae (LBV) to be $\sim10^{6.7}L_{\sun}$ if the star is at $2.3$ kpc.
For $\eta$ Carinae, we adopted the effective temperature of $\log{(T_{eff})}=4.128$ from the color--temperature relation
and the luminosity of $8\times10^6 L_{\sun}$ for the case of $d = 2.9$ kpc.
For HD 93162 (WN7), we adopted the effective temperature of $\log{(T_{eff})}=4.491$ and the luminosity of $10^6 L_{\sun}$ from \citet{cr95}.
\placefigure{hr}

\subsection{The Initial Mass Function\label{IMF}}
The H-R diagram with several evolutionary tracks is shown in Figure~\ref{hr}.
In the figure, most stars with $\log(T_{eff}) > 4.0$ follow a well-defined sequence near the ZAMS.
To determine the mass of individual stars, we adopted three different evolution models ---(i) the non-rotating evolution models of \citet{sc92}, 
(ii) the rotating models with the initial rotational velocity of $~330$ km s$^{-1}$ of \citet{brt11} for 5--60$M_\sun$ and the non-rotating models of \citet{sc92} for the other stars, 
and (iii) the rotating models of \citet{ek11}.
Because the mass of LBV and WN7 stars are very uncertain,
we assumed 75 $M_{\sun}$ \citep{ga08} and 100 $M_{\sun}$ for HD 93162 (WN7) and $\eta$ Carinae, respectively.
Although both stars seem to be binary systems,
we only considered the masses of the primaries because photometry alone cannot take into account the secondary of a binary system.
We also assumed the upper mass limit to be 100 $M_{\sun}$.
There are two more stars [HD 93129A (O2If$^*$) and HD 93250 (O3.5V((f+)))] with even higher mass. These two stars are brighter than the evolutionary track of a 120 $M_\sun$ star and suggests that the mass of the evolved stars (HD 93162 and $\eta$ Carinae) may be more massive than the upper mass limit (100 $M_{\sun}$, see also \citet{cr10}).
\placefigure{imf}

We calculated the IMF $\xi(\equiv N/\Delta\log m/$area) for the case of $\Delta \log m =0.2$.
The error bars are based on Poisson noise.
To minimize the binning effect, we also calculated the IMF by shifting 0.1 in $\log m$ and then calculated the slope of the IMF $\Gamma (\equiv d\log{\xi} /d\log m)$.
The slope of the IMF ($\Gamma$) calculated for members and candidates $-1.3\pm0.1$ (standard error) for Tr 14, $-1.3\pm0.1$ for Tr 16, and $-1.4\pm0.1$
for all observed regions down to our observational limit ($\log m \geqq0.2$, Table~\ref{tab_imf} case (1)).

Figure~\ref{imf} shows the IMF of Tr 14, Tr 16, and all observed area.
In the figure, the IMF calculated for members and candidates is marked as filled circles
and that calculated for members-only is marked as open circles.
All IMFs are the same as for Tr 14.
As Tr 14 is a compact cluster and most members are assembled in a small area, we can expect the contamination of field stars to be very low. On the other hand, the IMF of members and candidates of Tr 16 is slightly steeper than the IMF calculated for the members-only.
The difference is only apparent for $\log m < 0.6$.
From the figure, we can see that the IMF for members and candidates and the IMF for members-only differ somewhat for $\log m < 0.6$.
That is caused by the observational limit of the proper motion study \citep{cu93}.
Therefore, the IMF calculated for members including the candidates should be considered as an upper limit of the IMF of the $\eta$ Carina nebula region (Table~\ref{tab_imf} case (1)) 
and the IMF of the members-only can be considered as a lower limit (Table~\ref{tab_imf} case (2)).

If all stars in the PMS locus, regardless of membership selection criteria, are used in the calculation,
the IMF may be overestimated due to field star contamination.
To subtract the contribution of field stars, we should choose a control region representative of the field population toward the $\eta$ Carina nebula.
However, the choice of the field region is very difficult, especially for open clusters in the galactic plane.
If we choose a region far from the nebula, the background population will be counted more than the nebula region.
On the other hand, the region near the $\eta$ Carina nebula may represent both foreground and background populations,
but faint halo members of the cluster may also be subtracted as part of the ``foreground population".

In order to subtract the field contribution and compare the IMF before and after the subtraction,
we chose a small region in the NW part of the observed region.
The region is shown in Figure~\ref{plot} ($\Delta$ R.A.$<-13\farcm5$, $\Delta$ decl.$>0\farcm5$).
The location of the region coincides with the giant molecular cloud Car I \citep{br03}.
\citet{sm08} showed that the bright arc in H$\alpha$ is well matched with the curved edges of both $^{13}$CO and 3.3 $\mu$m PAH emission.
These facts allow us to expect that star formation has not yet propagated into this region,
and we can expect that no members of the $\eta$ Carina nebula exist in the region.
In addition, as the giant molecular cloud is very close to the $\eta$ Carina nebula, stars found in this region
should well represent the population of foreground stars.

We re-calculated the slope of the IMF, $\Gamma$, after subtraction of the field contribution down to our observational limit (Table~\ref{tab_imf} case (3)).
However, as there was no significant difference from the slopes of the IMF in Table~\ref{tab_imf} we conclude that the slope of IMF, $\Gamma$, is practically the same within the errors, with and without subtraction of the field contribution.
The field contamination in the observed region has no serious influence on the slope of the IMF.

The IMF of the $\eta$ Carina nebula is very similar to that of \object{NGC 6231} ($\Gamma =-1.2\pm0.4$; \citealt{su98}),
but is slightly shallower than the IMF of the solar neighborhood, the Pleiades, or NGC 2264 ($\Gamma \approx -1.7$; see Figure 6 of \citet{su04}).

\subsection{Age and Age Spread\label{age}}
The presence of the LBV ($\eta$ Carinae), WN7 (HD 93162), and several O3 stars indicates that
both Tr 14 and Tr 16 are very young clusters.
Most previous estimates of age and age spread were based on stellar evolutionary models.
\citet{va96} estimated 1.5 Myr for the most massive stars in Tr 14.
\citet{de01} estimated the age of both Tr 14 and Tr 16 to be 2 -- 3 Myr with an age spread of 5 -- 6 Myr for massive stars and 10 Myr for low-mass PMS stars.
\citet{ta03} estimated 5 Myr for Tr 14 and 3 -- 6 Myr for Tr 16.
Most recently, \citet{ca04} estimated 2 Myr for Tr 14 and 5 Myr for Tr 16.

\begin {figure}
\plotone{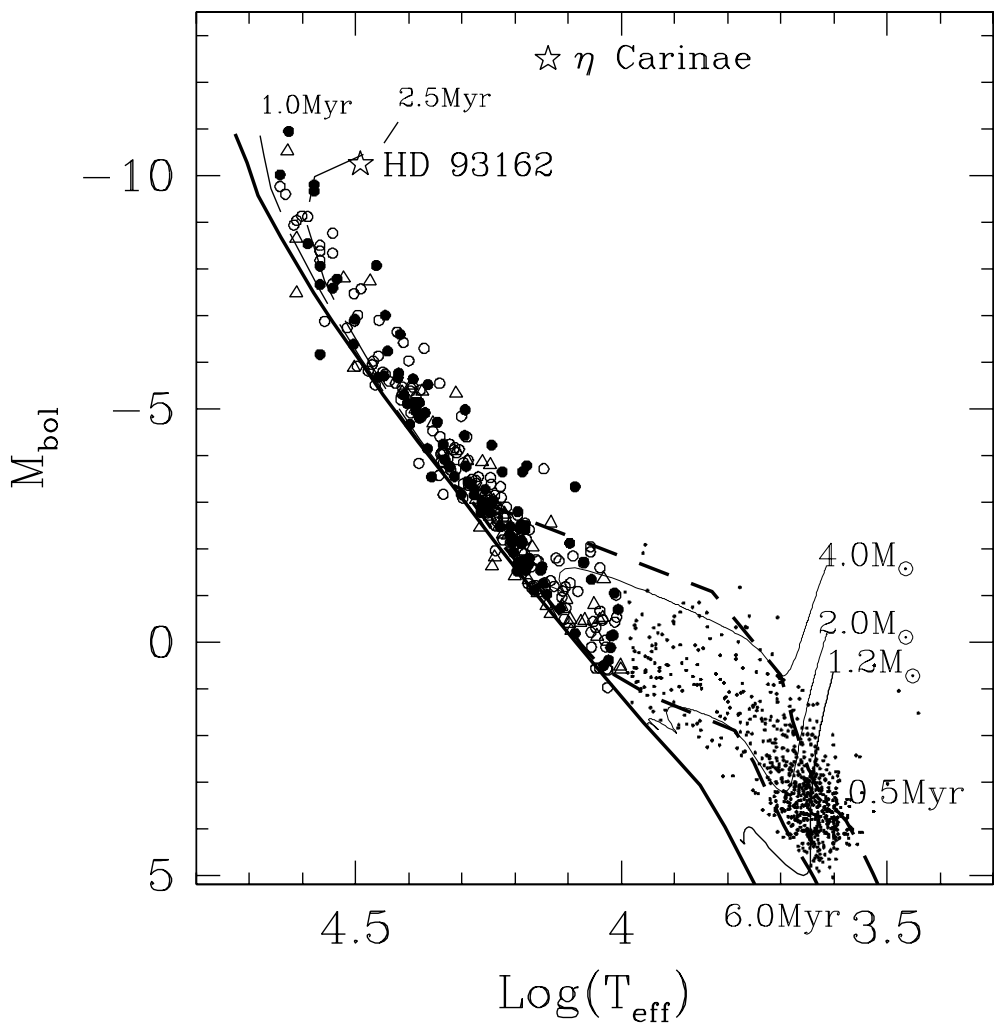}
\caption{
H-R diagram of members of the two clusters.
Dots, stars, filled circles, open circles, and open triangles represent PMS members, evolved stars (WN7, LBV), and
MS members of Tr 14, Tr 16, and other area in the observed region, respectively.
The thick and thin solid lines represent the ZAMS relation of \citet{sc92} and the PMS evolutionary tracks of \citet{si00}.
The thick and thin dashed lines represent the isochrones interpolated from the PMS evolutionary tracks and the MS and post-MS evolutionary models.
\label{hr}}
\end{figure}

\begin {figure}
\plotone{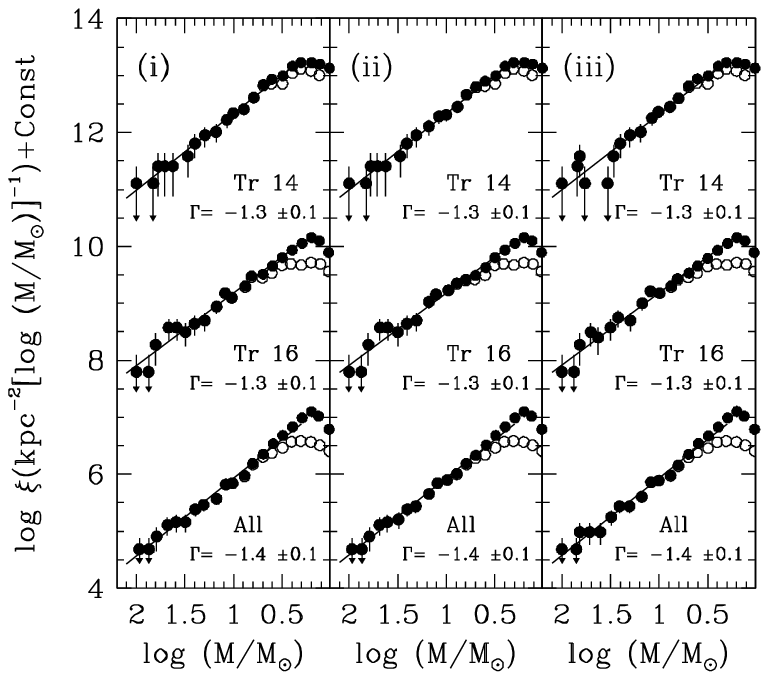}
\caption{
IMF of Tr 14, Tr 16, and all stars in the observed region.
Filled circles and open circles represent, respectively, the IMF for members and candidates (Table~\ref{tab_imf} case (2)), and for members-only (Table~\ref{tab_imf} case (1)).
The error bars are based on $\sqrt{N}$.\newline
(i) The IMF calculated using the non-rotating models of \citet{sc92}.\newline
(ii) The IMF calculated using the rotating models with the initial rotational 
velocity of ~330 km s$^{-1}$ from \citet{brt11} for 5--60$M_{\sun}$ and the non-rotating
models of \citet{sc92} for the other stars.\newline
(iii) The IMF calculated using the rotating models of \citet{ek11} for all stars.
\label{imf}}
\end{figure}

In Figure~\ref{hr}, two isochrones interpolated from the stellar evolution models of \citet{sc92} are superimposed.
HD 93128 (O3.5V((f+))) and HD 93129A (O2If$^*$) in Tr 14 and HD 93205 (O3.5V((f+))), HD 303308 (O4V((f+))) and ID 2022 (=Y398, O3-O4If) in Tr 16
are very close to the ZAMS and fit well to the 1 Myr isochrone, while HD 93162\footnote{
For HD 93162, \citet{ha06} obtained an effective temperature of 50,000K and luminosity of $10^{6.55} L_{\sun}$ for the case of $V_0-M_V=12.55$.   
If the effective temperature of \citet{ha06} and the luminosity corrected for the case of $V_0-M_V=12.3$ are applied, HD 93162 is much closer to the 1 Myr isochrone in Figure~\ref{hr}.}
 (WN7) in Tr 16 is well fitted to the 2.5 Myr isochrone. 
Therefore, the age of massive MS stars in both clusters is about 1 Myr.
The age of evolved stars in Tr 16 is about 2.5 Myr but we cannot estimate the upper limit of the age of Tr 14 because there is no evolved star in the cluster.

\begin {figure}
\plotone{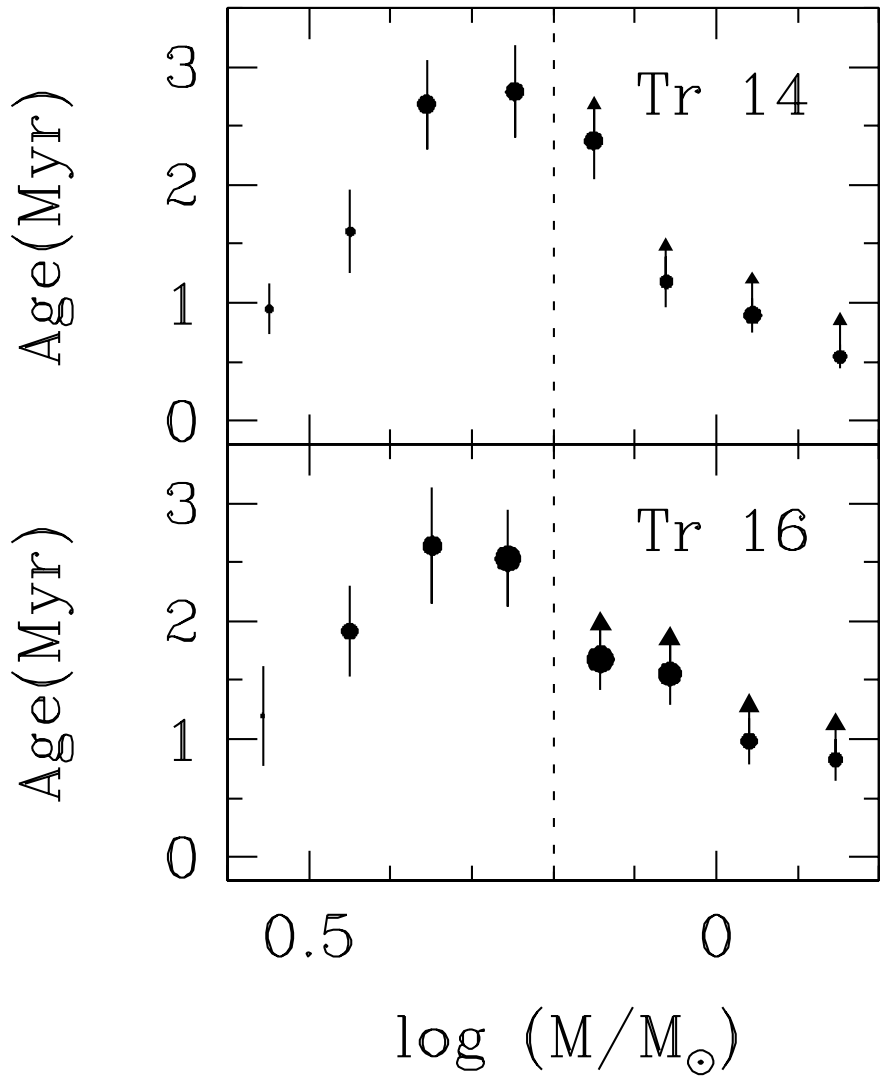}
\caption{
Age distribution of PMS members in Tr 14 and Tr 16. The size of dots is proportional to the number of PMS stars used in statistics.
The dashed line represents the completeness limit of this work ($\log m =0.2$).
For low-mass PMS stars, only young bright PMS members were observed and the age is therefore underestimated.
For high-mass PMS stars, PMS evolutionary models overestimate the age (see Section~\ref{age}).
\label{age_pms}}
\end{figure}

Figure~\ref{age_pms} shows the age distribution of PMS stars in the clusters.
The average age of PMS members in both clusters are distributed between 1 Myr and 3 Myr.
But, the age of the lower mass ($\log m <0.2$) stars seems to be older than 1 Myr,
because the more luminous PMS stars (thus younger PMS stars) are preferentially detected due to our observational limit.
On the other hand, the age of the massive PMS stars ($\log m \sim 0.3$) is about 3 Myr.
These massive PMS stars in both clusters appear to still be in the Kelvin--Helmholtz contraction phase.
\citet{su97,su04} suggested that the age of PMS stars in the Kelvin--Helmholtz contraction phase could be overestimated.
Later \citet{ha03} discussed the age of PMS stars determined from PMS models.

It is not easy to determine the age difference between these clusters
because the age distribution of both clusters seems to be very similar.
In order to estimate the difference in ages of the PMS stars in the clusters, deeper photometry down to sub-solar-mass stars is required.
\placefigure{age_pms}

\subsection{Is Cr 232 an independent cluster?}
It is uncertain whether Collinder 232 (Cr 232), a poor young cluster with a small number of blue stars, located to the eastern side of Tr 14  is a separate cluster or an extended part of Tr 14.
There are a few very early-type stars, including HD 93250 (O3.5V((f+))) and HD 303311 (O5V), that are thought to be members of Cr 232 together with other proper motion members, but this simply suggests that the stars are at the same distance and have similar kinematic properties to Tr 14 and Tr 16. The existence of O3.5V((f+)) stars in Cr 232 (HD 93250) and Tr 14 (HD 93128) also implies
that the two clusters are indistinguishable in age.
Previous studies have argued that Cr 232 may not be an independent cluster.
\citet{ta03} concluded that Cr 232 appears not to be a real cluster (see Figure 13 of \citet{ta03}).
\citet{ca04} also suggested that Cr 232 could be an extended part of Tr 14 rather than an independent cluster.
We agree with these conclusions and have treated Cr 232 as an outer region of Tr 14 and Tr 16 rather than a separate cluster.

\section{Summary\label{summ}}
We have presented new $UBVI_c$ photometric data down to $V=19$ for Tr 14 and Tr 16 obtained with the 1m telescope at Siding Spring Observatory and a SITe 2k CCD camera.
We selected the members and candidates using the X-ray source catalog from {\it Chandra} X-ray observations provided by \citet{al08}
and the CCCP X-ray point-source catalog provided by \citet{br11}, the
2MASS point-source catalog, and the membership probability from the proper motion study of \citet{cu93}.

We derived an abnormal reddening law in the $\eta$ Carina nebula and determined the foreground reddening to be $E(B-V)_{fg}=0.36\pm0.04$ mag.
An abnormal total-to-selective extinction ratio in the nebula, $R_{V,cl}=4.4\pm0.2$, was determined using 141 early-type members.
We derived the same distance modulus of $V_0-M_V=12.3\pm0.2$ mag ($d=2.9 \pm0.3$ kpc) for both Tr 14 and Tr 16,
and concluded that the clusters are at the same distance.

We derived the IMF and calculated the slope $\Gamma=-1.3\pm0.1$ for the $\eta$ Carina nebula region down to our observational limit ($\log m \geqq 0.2$).
The slope of the IMF, $\Gamma$, of the $\eta$ Carina nebula is very similar to that of NGC 6231
and is slightly shallower than that of NGC 2264, the Solar neighborhood, and the Pleiades.

We estimated the age of MS stars in both clusters to be about 1 Myr from the comparison between the theoretical evolution models
and the H-R diagram of the cluster.
The more evolved star HD 93162 in Tr 16 is matched well with the isochrone of age 2.5 Myr.
We have also estimated the age of the PMS stars in the $\eta$ Carina nebula to be 1 -- 3 Myr.

\acknowledgments
The authers thank the refree, D. Gies, for many insightful comments.
This work is supported by a National Research Foundation of Korea (NRF) grant funded by the Korea Government (MEST) (Grant No. 20110114136).

\clearpage

\clearpage

\begin{deluxetable}{rcccccccccccccccccccccccccccccccll}
\rotate
\tabletypesize{\scriptsize}
\tablewidth{0pt}
\tablecaption{Photometric Data\label{tab1}\tablenotemark{1}}
\rotate
\tablehead{
\colhead{ID} & \colhead{$\alpha_{J2000}$}  & \colhead{$\delta_{J2000}$} &
\colhead{$V$} & \colhead{$U-B$} & \colhead{$B-V$} & \colhead{$V-I$} &
\colhead{$\epsilon_V$} & \colhead{$\epsilon_{U-B}$} & \colhead{$\epsilon_{B-V}$} &
\colhead{$\epsilon_{V-I}$} & \colhead{N$_{obs}$} & \colhead{P$_\mu$\tablenotemark{1}} &
\colhead{remark\tablenotemark{2}} & \colhead{2MASS }}
\startdata
 1655&10 44 00.5&-59 33 03.9& 14.775&  0.271&  0.464&  0.560&  0.007&  0.021&  0.012&  0.024&6 2 6 6&  & X &10440043-5933030\\
 1656&10 44 00.5&-59 32 42.8& 18.538&\nodata&\nodata&  2.267&  0.026&\nodata&\nodata&  0.053&4 0 0 4&  & X &10440047-5932431\\
 1657&10 44 00.6&-59 48 54.6& 16.546&\nodata&  1.596&  2.435&  0.004&\nodata&  0.009&  0.008&6 0 5 6&  &   &10440062-5948544\\
 1658&10 44 00.6&-59 34 04.3& 13.834&  0.154&  0.483&  0.814&  0.002&  0.009&  0.006&  0.008&6 2 6 6&49&  I&10440061-5934043\\
 1659&10 44 00.6&-59 31 52.3& 13.415& -0.269&  0.287&  0.464&  0.007&  0.015&  0.011&  0.010&6 2 6 6&58&   &10440063-5931524\\
 1660&10 44 00.6&-59 32 07.0& 17.758&\nodata&  1.056&  1.147&  0.014&\nodata&  0.024&  0.022&4 0 4 4&  &   &                \\
 1661&10 44 00.7&-59 34 17.3& 15.053&  0.467&  1.097&  1.496&  0.002&  0.037&  0.011&  0.004&6 1 6 6&61&   &10440067-5934173\\
 1662&10 44 00.7&-59 32 33.5& 14.957&  0.177&  0.681&  0.740&  0.008&  0.046&  0.012&  0.020&6 2 6 6&  &   &10440072-5932334\\
 1663&10 44 00.7&-59 49 51.8& 12.495& -0.283&  0.161&  0.257&  0.007&  0.016&  0.018&  0.008&2 2 2 2&  &   &10440071-5949518\\
 1664&10 44 00.8&-59 45 03.8& 13.951&  1.178&  1.320&  1.421&  0.002&  0.023&  0.005&  0.004&6 1 6 6& 0&   &10440079-5945038\\
 1665&10 44 00.8&-59 34 38.9& 16.631&\nodata&  1.519&  1.976&  0.014&\nodata&  0.022&  0.019&6 0 5 6&  & X &10440081-5934390\\
 1666&10 44 00.8&-59 37 55.4& 18.387&\nodata&  1.070&  1.726&  0.019&\nodata&  0.071&  0.025&4 0 1 4&  &   &10440084-5937556\\
 1667&10 44 00.8&-59 48 59.4& 15.895&  0.767&  0.661&  1.282&  0.011&  0.038&  0.013&  0.032&6 1 6 3&  &D  &10440080-5948592\\
 1668&10 44 00.9&-59 33 51.1& 17.605&\nodata&  1.626&  2.194&  0.008&\nodata&  0.058&  0.010&4 0 1 4&  & X &10440091-5933512\\
 1669&10 44 00.9&-59 48 01.1& 18.911&\nodata&\nodata&  1.657&  0.021&\nodata&\nodata&  0.026&4 0 0 4&  &   &10440091-5948011\\
 1670&10 44 00.9&-59 35 45.7& 10.670& -0.625&  0.380&  0.712&  0.009&  0.012&  0.012&  0.012&4 1 4 4&96& X &10440093-5935458\\
 1671&10 44 00.9&-59 47 12.8& 16.048&  0.278&  0.825&  0.986&  0.004&  0.036&  0.007&  0.006&6 1 6 6&  &   &10440093-5947126\\
\enddata
\tablecomments{
Table \ref{tab1} is published in its entirety in the
electronic edition.  A portion is shown here for guidance regarding its form and content.}
\tablenotetext{1}{Membership probability ($P_\mu$ in percent) from proper motion study \citep{cu93}.}
\tablenotetext{2}{
D - The PSF of the object shows a convolution of two PSFs, but the object is
measured as a single star for the sake of photometric error. X - X-ray emission stars. I - NIR excess stars.
}
\end{deluxetable}
\clearpage

\begin{deluxetable}{lrrrrrrrr}
\rotate
\tabletypesize{\scriptsize}
\tablewidth{0pt}
\tablecaption{Comparison of Photometry\label{tab2}}
\tablehead{
Paper          &
$\Delta V$\tablenotemark{1}     & N(m)\tablenotemark{2} &
$\Delta$(\bv)\tablenotemark{1}  & N(m)\tablenotemark{2} &
$\Delta$(\ub)\tablenotemark{1}  & N(m)\tablenotemark{2} &
$\Delta (V-I)\tablenotemark{1}$ & N(m)\tablenotemark{2} }
\startdata
\citet{fe63}&$-0.006\pm 0.124$&  8(1) &$ 0.026\pm 0.023$&  8(1) &$ 0.034\pm 0.029$&  8(1) & &\\
\citet{fe69}&$-0.006\pm 0.128$& 35(2) &$ 0.007\pm 0.027$& 32(5) &$ 0.014\pm 0.034$& 34(3) & &\\
\citet{fe73}&$-0.041\pm 0.072$& 60(5) &$ 0.018\pm 0.022$& 58(7) &$-0.012\pm 0.041$& 57(8) & &\\
\citet{he76}&$ 0.007\pm 0.064$& 11(2) &$-0.006\pm 0.022$& 12(1) &$-0.032\pm 0.046$& 13(0) & &\\
\citet{fe82}&$-0.022\pm 0.062$& 26(5) &$ 0.028\pm 0.035$& 28(3) &$ 0.026\pm 0.061$& 26(3) &$ 0.016\pm0.036$& 38(12)\\
\citet{mj93}&$-0.003\pm 0.132$&160(23)&$ 0.031\pm 0.114$&171(12)&$-0.033\pm 0.076$&127(56)& &\\
\citet{va96}&$-0.006\pm 0.022$& 56(8) &$ 0.030\pm 0.030$& 55(9) &$ 0.018\pm 0.051$& 57(6) &$ 0.007\pm0.031$& 42(14)\\
\citet{ta03}&$-0.001\pm 0.039$&345(44)&$-0.045\pm 0.072$&371(18)&$ 0.080\pm 0.087$&140(18)&$-0.042\pm0.063$&356(33)\\
\citet{ca04}&$-0.018\pm 0.030$&138(11)&$ 0.014\pm 0.037$&140(9) &$ 0.060\pm 0.069$&134(14)&$ 0.072\pm0.044$&177(18)\\
\enddata
\tablenotetext{1}{This - Others}
\tablenotetext{2}{N and m represent number of compared stars and excluded stars.}
\end{deluxetable}
\clearpage

\begin{deluxetable}{lccccl}
\tabletypesize{\scriptsize}
\tablewidth{0pt}
\tablecaption{Comparison of the IMF slope \label{tab_imf}}
\tablehead{Case & Mass Range & Tr 14 & Tr 16 & All Region & Note}
\startdata
(1) & ($\log m\geqq0.2$) & $-1.3\pm 0.1$ & $-1.3\pm 0.1$ & $-1.4\pm 0.1$ & Member + candidate \\
(2) & ($\log m\geqq0.2$) & $-1.3\pm 0.1$ & $-1.1\pm 0.1$ & $-1.2\pm 0.1$ & Member only \\
(3) & ($\log m\geqq0.2$) & $-1.3\pm 0.1$ & $-1.2\pm 0.1$ & $-1.3\pm 0.1$ & Member + candidate + field correction \\
\enddata
\end{deluxetable}
\clearpage

\end{document}